# Equilibrium Points and Orbits around Asteroid with the Full Gravitational Potential Caused by the 3D Irregular Shape

Yu Jiang[1, 2]

1. State Key Laboratory of Astronautic Dynamics, Xi'an Satellite Control Center, Xi'an 710043, China
2. School of Aerospace Engineering, Tsinghua University, Beijing 100084, China

Y. Jiang (✉) e-mail: jiangyu_xian_china@163.com (corresponding author)

**Abstract**. We investigate the equilibrium points and orbits around asteroid 1333 Cevenola by considering the full gravitational potential caused by the 3D irregular shape. The gravitational potential and effective potential of asteroid 1333 Cevenola are calculated. The zero-velocity curves for a massless particle orbiting in the gravitational environment have been discussed. The linearized dynamic equation, the characteristic equation, and the conserved quantity of the equilibria for the large-size-ratio binary asteroid system have been derived. It is found that there are totally five equilibrium points close to 1333 Cevenola. The topological cases of the outside equilibrium points have a staggered distribution. The simulation of orbits in the full gravitational potential caused by the 3D irregular shape of 1333 Cevenola shows that the moonlet's orbit is more likely to be stable if the orbit inclination is small.

## 1 Introductions

The orbital mechanics in the full complex gravitational field of the asteroid caused by the 3D irregular shape has became an important topic in recent years. The study of the orbital mechanics around asteroids is useful for the orbital design of the deep space missions to asteroids. In addition, the discovery of moonlets of asteroids also makes the study of the orbital mechanics around asteroids useful for understanding the formation and evolution of binary asteroid systems.

Previous studies have used simplified models to model the shape and gravitational field of asteroids. Hirabayashi et al. [1] used a uniformly rotating system composed of two masses connected by a massless rod to model the shape and gravitational field of asteroid $2000EB_{14}$, and found that there are three equilibrium points in the x-axis (long axis) of the body-fixed frame of the asteroid, then they analyzed the stability of these equilibrium points. Zeng et al. [2] used the same model with Hirabayashi et al.[1] to model the shape and the gravitational field caused by the shape of several different minor celestial bodies. They called the model as the rotating mass dipole. These minor celestial bodies include asteroids 216 Kleopatra, 951 Gaspra, 1620 Geographos, 1996 HW1, 2063 Bacchus, and 25143 Itokawa, as well as the comet 103P/Hartley-2. They calculated the positions and eigenvalues of equilibrium points around these minor celestial bodies. Yang et al. [3,4] also applied the rotating mass dipole to model the gravitational field of asteroids 216 Kleopatra, 951 Gaspra, 1620 Geographos, 1996 HW1, 2063 Bacchus, and 25143 Itokawa. Venditti et al. [5] used the parallelepipeds to model the gravitational potential of non-spherical bodies and calculated orbits around the bodies. Feng et al. [6] applied the ellipsoid-sphere to model the shape and gravitational potential of contact binary asteroid 1996 HW1, and calculated the position of equilibrium points and orbits around equilibrium points.

Although using the simplified models to model the shape and gravitational field of asteroids has faster compute speed with computer (Przemysław et al. [7]), the positions of equilibrium points may have significant errors. Jiang et al. [8] calculated the positions and eigenvalues of asteroids 216 Kleopatra, 1620 Geographos, 4769Castalia, and 6489 Golevka with considering the full gravitational potential caused by the asteroids' 3D irregular shapes. Wang et al. [9] calculated the positions of several irregular minor celestial bodies which also considered the full gravitational potential caused by the 3D irregular shapes, the results include the asteroids 216 Kleopatra, 951 Gaspra, and 1620 Geographos, the contact binary asteroids 1996 HW1, 2063 Bacchus, 4769 Castalia, 25143 Itokawa, as well as the comets 103P/Hartley-2. Using the simplified models, the equilibrium points of the asteroids are calculated in the xy plane of the body [1-6]. However, all the equilibrium points of the known asteroids are out-of-plane equilibrium points [8-10] because of the irregular shape of the body.

Besides, the simplified models may cause the loss of some important characteristics in the gravitational potential of the uniformly rotating irregular shaped bodies. These characteristics include the number of equilibrium points (Yu and Baoyin [11]; Wang et al. [9]), the stability and topological cases of equilibrium points (Jiang et al. [8]; Wang et al. [9]), the topological structure of zero-velocity curves (Yu and Baoyin [11]; Chanut et al. [12,13]), the number of families of periodic orbits (Yu and baoyin [14]; Jiang and Baoyin [15]), the impact of the orbit on the surfaces (Tardivel et al. [16]; Yu and Baoyin [17]; Jiang et al. [18]), the transfer of orbital motions to surface motions (Jiang et al. [18]; Wal and Scheeres [19]), etc.

In this paper, we study the dynamical characteristics in the gravitational potential of the asteroid 1333 Cevenola, which is a uniformly rotating irregular shaped minor celestial body. The bulk density for asteroid 1333 Cevenola is 1.6 g·cm$^{-3}$ (Johnston [20,21]), and the rotational period is 4.88h (Warner [22]). Using the polyhedron model, we calculated the size of asteroid 1333 Cevenola to be $18.971 \times 12.051 \times 10.283$km. The mass is calculated to be $1.6 \times 10^{15}$kg. The full gravitational potential caused by the 3D irregular shape is also calculated. Ni et al. [23] analyzed the size of the perturbation from the solar gravity relative to the gravity of the asteroids, and indicated that the solar gravity is a high-order small quantity relative to the asteroidal gravity when the particle is close to the asteroid. Thus we neglect the perturbation acting on the particle from the solar gravity. We investigated the shape model of 1333 Cevenola using the polyhedron model method with 1018 vertexes, 3048 edges, as well as 2032 faces. The linearized dynamic equation and the characteristic equation for the large-size-ratio binary asteroid system have been derived. In addition, a conserved quantity of the equilibria for the large-size-ratio binary asteroid has been found. The full gravitational potential and effective potential of 1333 Cevenola are analyzed. The zero-velocity curves for a massless particle in the full gravitational environment of the asteroid 1333 Cevenola have been investigated, and it is found that there are five equilibrium points close to 1333 Cevenola. We also computed two different orbits in the full gravitational potential caused by the 3D irregular shape of 1333 Cevenola, the results state that the moonlet's orbit is more likely to be stable if the orbit inclination is small. The results are useful to study the equilibrium points and orbits for a spacecraft moving around the large-size-ratio binary asteroid system, and also helpful for understanding the relative equilibrium of the synchronous binary asteroids.

## 2 Motion Equation, Polyhedral Model, and Gravitational Potential

Relative to the body-fixed coordinate frame, the motion equation (Jiang and Baoyin [24]) of the particle can be expressed by

$$\begin{cases} \ddot{x} + \dot{\omega}_y z - \dot{\omega}_z y + 2\omega_y \dot{z} - 2\omega_z \dot{y} + \omega_x \omega_y y - \omega_y^2 x - \omega_z^2 x + \omega_z \omega_x z + \dfrac{\partial U}{\partial x} = 0 \\ \ddot{y} + \dot{\omega}_z x - \dot{\omega}_x z + 2\omega_z \dot{x} - 2\omega_x \dot{z} + \omega_y \omega_z z - \omega_z^2 y - \omega_x^2 y + \omega_x \omega_y x + \dfrac{\partial U}{\partial y} = 0 \\ \ddot{z} + \dot{\omega}_x y - \dot{\omega}_y x + 2\omega_x \dot{y} - 2\omega_y \dot{x} + \omega_x \omega_z x - \omega_x^2 z - \omega_y^2 z + \omega_y \omega_z y + \dfrac{\partial U}{\partial z} = 0 \end{cases} \quad (1)$$

Assuming that the asteroid rotates uniformly, then the above equation can be reduced by

$$\begin{cases} \ddot{x} + 2\omega_y \dot{z} - 2\omega_z \dot{y} + \omega_x \omega_y y - \omega_y^2 x - \omega_z^2 x + \omega_z \omega_x z + \dfrac{\partial U}{\partial x} = 0 \\ \ddot{y} + 2\omega_z \dot{x} - 2\omega_x \dot{z} + \omega_y \omega_z z - \omega_z^2 y - \omega_x^2 y + \omega_x \omega_y x + \dfrac{\partial U}{\partial y} = 0 , \\ \ddot{z} + 2\omega_x \dot{y} - 2\omega_y \dot{x} + \omega_x \omega_z x - \omega_x^2 z - \omega_y^2 z + \omega_y \omega_z y + \dfrac{\partial U}{\partial z} = 0 \end{cases} \quad (2)$$

where $\nabla U = \begin{bmatrix} \dfrac{\partial U}{\partial x} & \dfrac{\partial U}{\partial y} & \dfrac{\partial U}{\partial z} \end{bmatrix}^T$ is the gravitational force acceleration.

Define the effective potential as

$$V(\mathbf{r}) = -\frac{1}{2}(\boldsymbol{\omega} \times \mathbf{r}) \cdot (\boldsymbol{\omega} \times \mathbf{r}) + U(\mathbf{r}). \quad (3)$$

Where $\mathbf{r} = \begin{bmatrix} x & y & z \end{bmatrix}^T$ and $\boldsymbol{\omega} = \omega \mathbf{e}_z$. We assume that the asteroid rotates around its largest moment of inertia, and that the z-axis of the body-fixed frame points to the largest moment of inertia of the asteroid. Let the x-axis points to the smallest moment of inertia, and y-axis points to the intermediate moment of inertia. Then the motion equation of the particle can be simplified into

$$\begin{cases} \ddot{x} - \dot{\omega}y - 2\omega\dot{y} + \dfrac{\partial V}{\partial x} = 0 \\ \ddot{y} + \dot{\omega}x + 2\omega\dot{x} + \dfrac{\partial V}{\partial y} = 0 \\ \ddot{z} + \dfrac{\partial V}{\partial z} = 0 \end{cases} \quad (4)$$

The Hamilton function $H$ and the mechanical energy $E$ of the particle is

$$H = \frac{1}{2}\dot{\mathbf{r}} \cdot \dot{\mathbf{r}} + V(\mathbf{r}). \quad (5)$$

$$E = \frac{1}{2}(\dot{\mathbf{r}} + \boldsymbol{\omega} \times \mathbf{r}) \cdot (\dot{\mathbf{r}} + \boldsymbol{\omega} \times \mathbf{r}) + U(\mathbf{r}). \quad (6)$$

Using Eq. (5), one can calculate the zero-velocity curves (Yu and Baoyin [11]; Scheeres et al. [25]; Dullin and Worthington [26]) of the particle:

$$V(\mathbf{r}) = H. \quad (7)$$

The inequality $V(\mathbf{r}) > H$ corresponds to the forbidden region for the particle, and $V(\mathbf{r}) < H$ corresponds to the allowed region for the particle. The equation $V(\mathbf{r}) = H$ means the particle's velocity equals zero relative to the body of the asteroid.

The gravitational potential and force (Werner and Scheeres [27]) of the body can be calculated using the polyhedron model. All the surface configurations of the polyhedron model presented in Werner and Scheeres [27] are triangles. Using the Euler formula

$$F_a + V_e - E_d = 2, \quad (9)$$

where $E_d$ means number of edges, $V_e$ means number of vertexes, $F_a$ means number of faces. Considering the triangles, we have

$$2E_d = 3F_a. \quad (10)$$

Thus, for the polyhedron model, we have

$$2V_e\text{-}4=F_a. \tag{11}$$

The polyhedron model of 1333 Cevenola (Hanus et al. [25]) here is built by 1018 vertexes, 3048 edges, and 2032 faces. Figure 1 illustrates the 3D polyhedron model of asteroid 1333 Cevenola viewed from different visual angles. From Figure 1, one can see that the shape of 1333 Cevenola is irregular, asymmetric, and convex. The projection of the shape in the equatorial plane, yz plane, and zx plane are different. Figure 2 invistigates the relative size of surface parameters of asteroid 1333 Cevenola, including surface height, surface gravitation, surface effective gravitation, as well as surface effective potential. The surface height is defined as the distance from the mass center to the mass point on the surface. The surface gravitation is $\nabla U = \begin{bmatrix} \dfrac{\partial U}{\partial x} & \dfrac{\partial U}{\partial y} & \dfrac{\partial U}{\partial z} \end{bmatrix}^{\mathrm{T}}$ on the surface. The surface effective gravitation is $\nabla V = \begin{bmatrix} \dfrac{\partial V}{\partial x} & \dfrac{\partial V}{\partial y} & \dfrac{\partial V}{\partial z} \end{bmatrix}^{\mathrm{T}}$ on the surface. From Figure 2, one can see that the surface parameters on different regions of the irregular body are quite different.

Figure 3 shows the gravitational potential of asteroid 1333 Cevenola in the xy, yz, and zx planes, respectively. Near different surface regions of asteroid 1333 Cevenola, the gravitational potential is quite different. The irregular shape's influence to the gravitational potential is obvious near the surface.

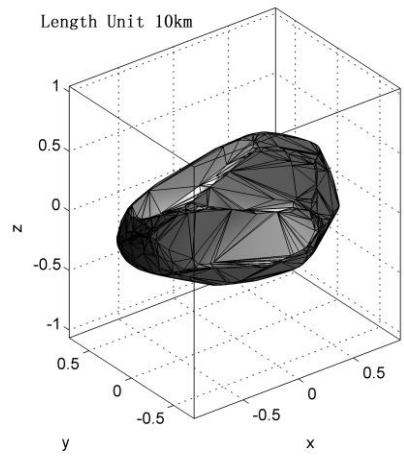

(a)

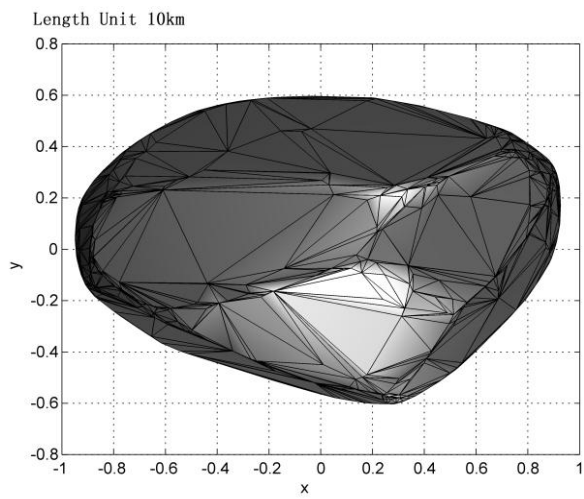

(b)

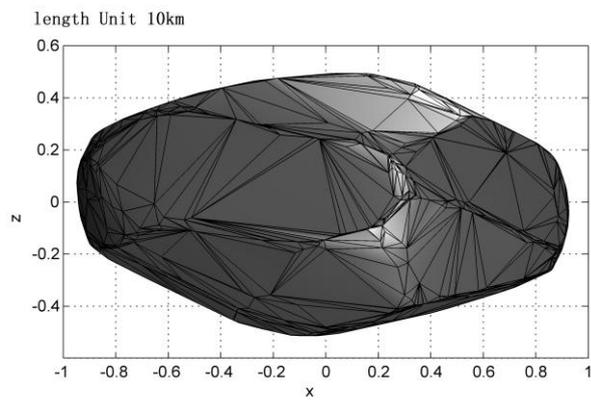

(c)

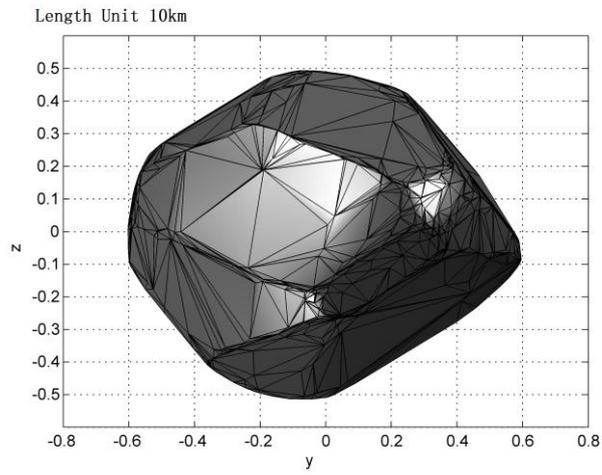

(d)

Figure 1 3D polyhedron model of asteroid 1333 Cevenola viewed from different visual angles. The shape is built with 2032 faces and 1018 vertices. (a) 3D View; (b) Viewed in xy plane; (c) Viewed in xz plane; (d) Viewed in yz plane.

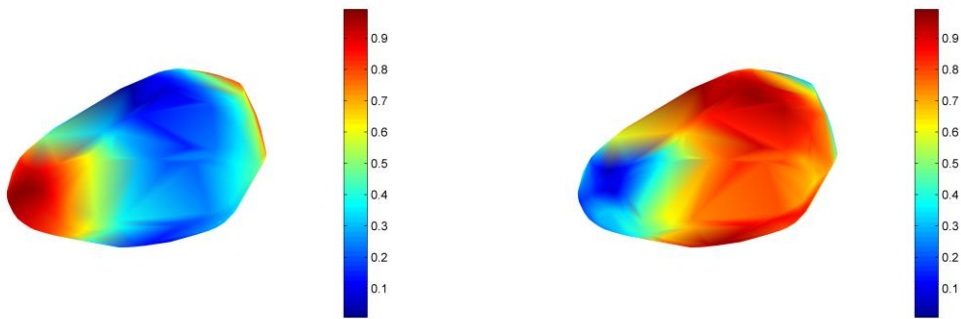

(a)                      (b)

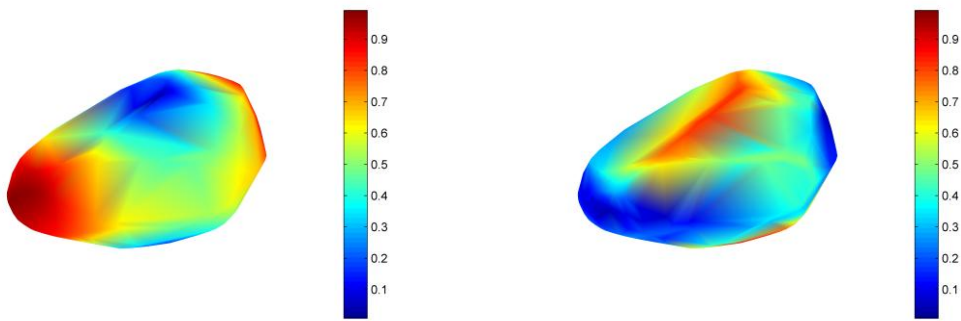

(c)                                    (d)

Figure 2 Relative size of surface parameters of asteroid 1333 Cevenola. (a) surface height; (b) surface gravitation; (c) surface effective gravitation; (d) surface effective potential.

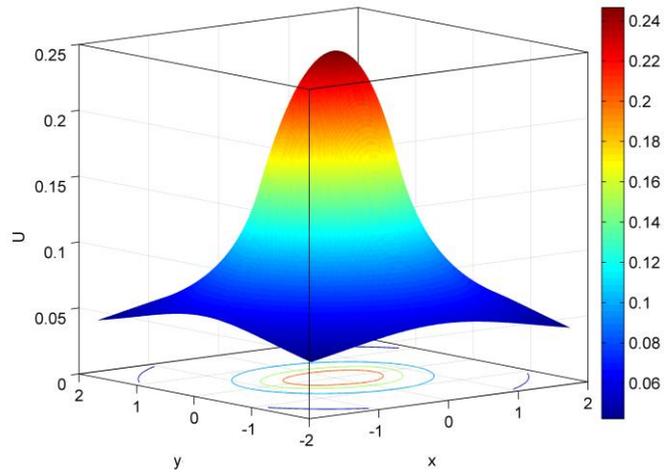

(a)

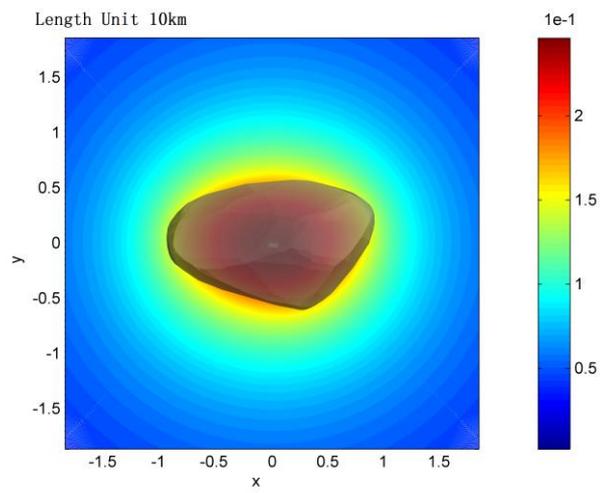

(b)

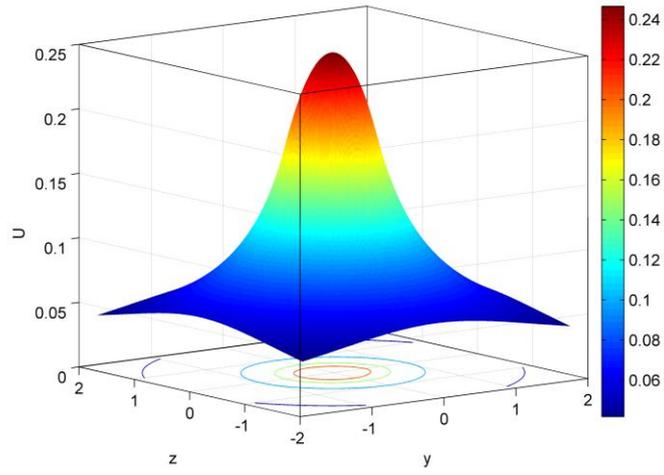

(c)

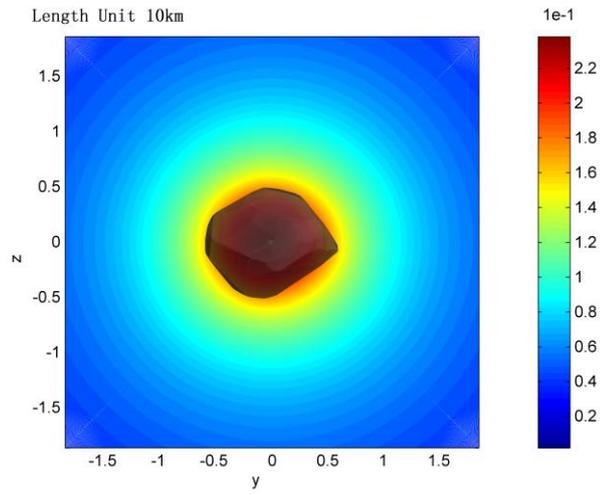

(d)

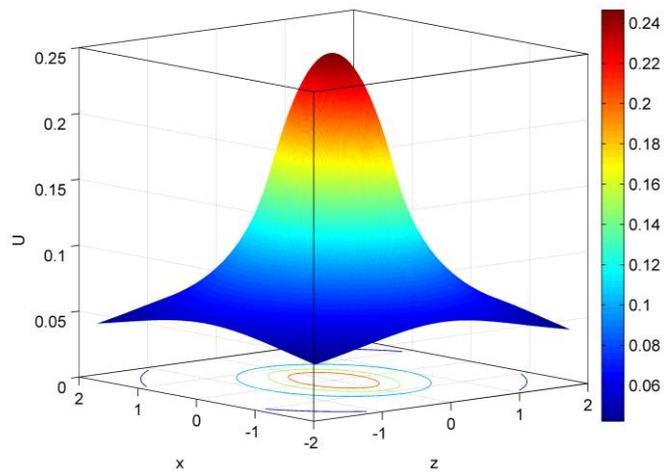

(e)

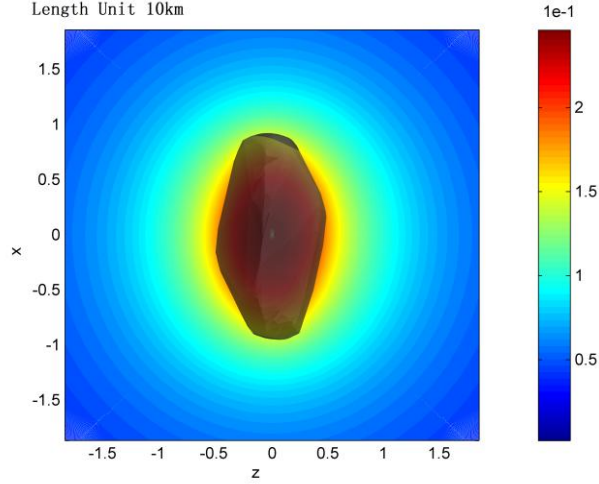

(f)

Figure 3 Gravitational potential of asteroid 1333 Cevenola in the xy, yz, and zx planes, respectively. The unit of the colour code is $10^2 m^2 \cdot s^{-2}$. The unit of the lengths in x, y, and z axes is 10 km. (a) 3D plot of the gravitational potential of asteroid 1333 Cevenola in the xy plane; (b) The shape and the contour plot of the gravitational potential of asteroid 1333 Cevenola in the xy plane; (c) 3D plot of the gravitational potential of asteroid 1333 Cevenola in the yz plane; (d) The shape and the contour plot of the gravitational potential of asteroid 1333 Cevenola in the yz plane; (e) 3D plot of the gravitational potential of asteroid 1333 Cevenola in the zx plane; (f) The shape and the contour plot of the gravitational potential of asteroid 1333 Cevenola in the zx plane.

**3 Effective Potential and equilibrium points**

In this section, we first derive the equilibria of the large-size-ratio binary asteroid system, and then we derive the equilibria of the primary. The equilibria of the primary is also the equilibria of a single asteroid.

**3.1 Equilibrium points of the large-size-ratio binary asteroid**

The large-size-ratio binary asteroid can be modeled by a highly irregular-shaped celestial body and a spherical body. The motion equation [28] can be written in the normalized frame as

$$\begin{cases} \ddot{\mathbf{r}} + 2\boldsymbol{\omega}\times\dot{\mathbf{r}} + \dot{\boldsymbol{\omega}}\times\mathbf{r} + \boldsymbol{\omega}\times(\boldsymbol{\omega}\times\mathbf{r}) = \dfrac{\partial u}{\partial \mathbf{r}} \\ \mathbf{I}\cdot\dot{\boldsymbol{\omega}} + \boldsymbol{\omega}\times\mathbf{I}\cdot\boldsymbol{\omega} = -\gamma\mathbf{r}\times\dfrac{\partial u}{\partial \mathbf{r}} \end{cases}, \qquad (12)$$

Where $u$ is the gravity potential of the irregular-shaped celestial body, $\mathbf{I}$ is the inertia dyad, and $\gamma$ is the coefficient calculated by the masses of these two bodies. The above model is also called the Sphere-Restricted Full 2-Body Problem (SRF2BP).

Let the time derivative equal zero, one can get the condition of the equilibrium state

$$\begin{cases} \boldsymbol{\omega}\times(\boldsymbol{\omega}\times\mathbf{r}) = \dfrac{\partial u}{\partial \mathbf{r}} \\ \boldsymbol{\omega}\times\mathbf{I}\cdot\boldsymbol{\omega} + \gamma\mathbf{r}\times\dfrac{\partial u}{\partial \mathbf{r}} = 0 \end{cases}. \qquad (13)$$

Denote

$$\begin{cases} \mathbf{g} = \boldsymbol{\omega}\times(\boldsymbol{\omega}\times\mathbf{r}) - \dfrac{\partial u}{\partial \mathbf{r}} = \begin{bmatrix} g_1 & g_2 & g_3 \end{bmatrix}^T \\ \mathbf{h} = \boldsymbol{\omega}\times\mathbf{I}\cdot\boldsymbol{\omega} + \gamma\mathbf{r}\times\dfrac{\partial u}{\partial \mathbf{r}} = \begin{bmatrix} h_1 & h_2 & h_3 \end{bmatrix}^T \end{cases}. \qquad (14)$$

Then the dynamic equation for the large-size-ratio binary asteroid reduces to

$$\begin{cases} \ddot{\mathbf{r}} + 2\boldsymbol{\omega}\times\dot{\mathbf{r}} + \dot{\boldsymbol{\omega}}\times\mathbf{r} + \mathbf{g} = 0 \\ \mathbf{I}\cdot\dot{\boldsymbol{\omega}} + \mathbf{h} = 0 \end{cases}. \qquad (15)$$

We now expand the dynamic equation at the equilibrium state. Denote s as the equilibrium state. Let

$$\begin{cases} \xi = x - x(s), \eta = y - y(s), \varsigma = z - z(s) \\ \alpha = \omega_x - \omega_x(s), \beta = \omega_y - \omega_y(s), \gamma = \omega_z - \omega_z(s) \end{cases}, \qquad (16)$$

and

$$g_{1x} = \left.\frac{\partial g_1}{\partial x}\right|_S, g_{1y} = \left.\frac{\partial g_1}{\partial y}\right|_S, g_{1z} = \left.\frac{\partial g_1}{\partial z}\right|_S. \tag{17}$$

Similarly, define $g_{2x}, g_{2y}, g_{2z}, g_{3x}, g_{3y}, g_{3z}, h_{1x}, h_{1y}, h_{1z}, h_{2x}, h_{2y}, h_{2z}, h_{3x}, h_{3y}, h_{3z}$, let

$\mathbf{I}^{-1} = \begin{bmatrix} \mathbf{i}_1 & \mathbf{i}_2 & \mathbf{i}_3 \end{bmatrix}$. Let $\mathbf{h}_x = [h_{1x}, h_{2x}, h_{3x}]^T$, $\mathbf{h}_y = [h_{1y}, h_{2y}, h_{3y}]^T$, and $\mathbf{h}_z = [h_{1z}, h_{2z}, h_{3z}]^T$.

Then the linearized dynamic equation at the equilibrium state can be given as

$$\begin{cases} \ddot{\xi} - 2\omega\dot{\eta} - \dot{\omega}\eta + g_{1x}\xi + g_{1y}\eta + g_{1z}\varsigma = 0 \\ \ddot{\eta} + 2\omega\dot{\xi} + \dot{\omega}\xi + g_{2x}\xi + g_{2y}\eta + g_{2z}\varsigma = 0 \\ \ddot{\varsigma} + g_{3x}\xi + g_{3y}\eta + g_{3z}\varsigma = 0 \\ \dot{\alpha} + (\mathbf{i}_1 \cdot \mathbf{h}_x)\alpha + (\mathbf{i}_1 \cdot \mathbf{h}_y)\beta + (\mathbf{i}_1 \cdot \mathbf{h}_z)\gamma = 0 \\ \dot{\beta} + (\mathbf{i}_2 \cdot \mathbf{h}_x)\alpha + (\mathbf{i}_2 \cdot \mathbf{h}_y)\beta + (\mathbf{i}_2 \cdot \mathbf{h}_z)\gamma = 0 \\ \dot{\gamma} + (\mathbf{i}_3 \cdot \mathbf{h}_x)\alpha + (\mathbf{i}_3 \cdot \mathbf{h}_y)\beta + (\mathbf{i}_3 \cdot \mathbf{h}_z)\gamma = 0 \end{cases} \tag{18}$$

It has 9 eigenvalues, the characteristic equation is

$$\begin{vmatrix} \lambda^2 + g_{1x} & -2\omega\lambda + (-\dot{\omega} + g_{1y}) & g_{1z} \\ 2\omega\lambda + (\dot{\omega} + g_{2x}) & \lambda^2 + g_{2y} & g_{2z} \\ g_{3x} & g_{3y} & \lambda^2 + g_{3z} \end{vmatrix} \cdot \begin{vmatrix} \lambda + (\mathbf{i}_1 \cdot \mathbf{h}_x) & (\mathbf{i}_1 \cdot \mathbf{h}_y) & (\mathbf{i}_1 \cdot \mathbf{h}_z) \\ (\mathbf{i}_2 \cdot \mathbf{h}_x) & \lambda + (\mathbf{i}_2 \cdot \mathbf{h}_y) & (\mathbf{i}_2 \cdot \mathbf{h}_z) \\ (\mathbf{i}_3 \cdot \mathbf{h}_x) & (\mathbf{i}_3 \cdot \mathbf{h}_y) & \lambda + (\mathbf{i}_3 \cdot \mathbf{h}_z) \end{vmatrix} = 0.$$

$$\tag{19}$$

i.e.

$$\begin{vmatrix} \lambda^2 + g_{1x} & -2\omega\lambda + (-\dot{\omega} + g_{1y}) & g_{1z} \\ 2\omega\lambda + (\dot{\omega} + g_{2x}) & \lambda^2 + g_{2y} & g_{2z} \\ g_{3x} & g_{3y} & \lambda^2 + g_{3z} \end{vmatrix} = 0, \tag{20}$$

or

$$\begin{vmatrix} \lambda + (\mathbf{i}_1 \cdot \mathbf{h}_x) & (\mathbf{i}_1 \cdot \mathbf{h}_y) & (\mathbf{i}_1 \cdot \mathbf{h}_z) \\ (\mathbf{i}_2 \cdot \mathbf{h}_x) & \lambda + (\mathbf{i}_2 \cdot \mathbf{h}_y) & (\mathbf{i}_2 \cdot \mathbf{h}_z) \\ (\mathbf{i}_3 \cdot \mathbf{h}_x) & (\mathbf{i}_3 \cdot \mathbf{h}_y) & \lambda + (\mathbf{i}_3 \cdot \mathbf{h}_z) \end{vmatrix} = 0. \tag{21}$$

Let

$$\mathbf{f} = \begin{bmatrix} \mathbf{g} \\ \mathbf{h} \end{bmatrix}. \tag{22}$$

Then the linearized dynamic equation can be written as

$$\begin{cases} \mathbf{M}\ddot{\boldsymbol{\rho}} + \mathbf{G}\dot{\boldsymbol{\rho}} + \mathbf{K}\boldsymbol{\rho} = 0 \\ \mathbf{M}\dot{\boldsymbol{\varepsilon}} + \mathbf{L}\boldsymbol{\varepsilon} = 0 \end{cases}. \tag{23}$$

Where $\boldsymbol{\rho} = \begin{bmatrix} \xi & \eta & \varsigma \end{bmatrix}^T$, $\mathbf{M}$ is a $3 \times 3$ unit matrix, $\mathbf{G} = \begin{pmatrix} 0 & -2\omega & 0 \\ 2\omega & 0 & 0 \\ 0 & 0 & 0 \end{pmatrix}$,

$$\mathbf{K} = \begin{bmatrix} g_{1x} & g_{1y} & g_{1z} \\ g_{2x} & g_{2y} & g_{2z} \\ g_{3x} & g_{3y} & g_{3z} \end{bmatrix} = \nabla \mathbf{g}(S),$$

$$\mathbf{L} = \begin{bmatrix} \lambda + (\mathbf{i}_1 \cdot \mathbf{h}_x) & (\mathbf{i}_1 \cdot \mathbf{h}_y) & (\mathbf{i}_1 \cdot \mathbf{h}_z) \\ (\mathbf{i}_2 \cdot \mathbf{h}_x) & \lambda + (\mathbf{i}_2 \cdot \mathbf{h}_y) & (\mathbf{i}_2 \cdot \mathbf{h}_z) \\ (\mathbf{i}_3 \cdot \mathbf{h}_x) & (\mathbf{i}_3 \cdot \mathbf{h}_y) & \lambda + (\mathbf{i}_3 \cdot \mathbf{h}_z) \end{bmatrix} = \mathbf{I}^{-1} \nabla \mathbf{h}(S).$$

Denote $\dot{\boldsymbol{\rho}} = \boldsymbol{\sigma}$ and $\boldsymbol{\Lambda} = \begin{bmatrix} \boldsymbol{\rho} & \boldsymbol{\sigma} & \boldsymbol{\varepsilon} \end{bmatrix}^T$, we have

$$\begin{bmatrix} \dot{\boldsymbol{\rho}} \\ \dot{\boldsymbol{\sigma}} \\ \dot{\boldsymbol{\varepsilon}} \end{bmatrix} = \begin{bmatrix} 0 & \mathbf{M} & 0 \\ -\mathbf{M}^{-1}\mathbf{K} & -\mathbf{M}^{-1}\mathbf{G} & 0 \\ 0 & 0 & \mathbf{M}^{-1}\mathbf{L} \end{bmatrix} \begin{bmatrix} \boldsymbol{\rho} \\ \boldsymbol{\sigma} \\ \boldsymbol{\varepsilon} \end{bmatrix} = \begin{bmatrix} 0 & \mathbf{M} & 0 \\ -\mathbf{K} & -\mathbf{G} & 0 \\ 0 & 0 & \mathbf{L} \end{bmatrix} \begin{bmatrix} \boldsymbol{\rho} \\ \boldsymbol{\sigma} \\ \boldsymbol{\varepsilon} \end{bmatrix}, \tag{24}$$

i.e.

$$\dot{\boldsymbol{\Lambda}} = \mathbf{A}\boldsymbol{\Lambda}. \tag{25}$$

Where $\mathbf{A} = \begin{bmatrix} 0 & \mathbf{M} & 0 \\ -\mathbf{K} & -\mathbf{G} & 0 \\ 0 & 0 & \mathbf{L} \end{bmatrix}$.

For the number of the equilibrium states [15,18, 29] of the SRF2BP, we have the following conserved quantity

$$\det \nabla \mathbf{f} = \det(\nabla \mathbf{g}(S)) \cdot \det(\mathbf{I}^{-1} \nabla \mathbf{h}(S)) = \det \mathbf{A} = (\det \mathbf{K}) \cdot (\det \mathbf{L}). \tag{26}$$

Then

$$\sum_j \text{sgn}(\det \nabla \mathbf{f}) = \sum_j \text{sgn}(\det \mathbf{A}(S_j)) = \sum_j \text{sgn}((\det \mathbf{K}) \cdot (\det \mathbf{L})) = \text{const}. \tag{27}$$

Here $S_j$ represents the j-th equilibrium state of the large-size-ratio binary asteroid.

### 3.2 Equilibrium points of the primary

From Eq. (5), one can see that for a relative equilibrium points, the velocity of the particle on the equilibrium points relative to the body-fixed frame is zero. Thus we have

$$H = V(\mathbf{r}) = \text{const.}, \tag{28}$$

and the positions of the equilibrium points satisfy

$$\frac{\partial V(\mathbf{r})}{\partial \mathbf{r}} = 0. \tag{29}$$

Thus we know that the positions of the relative equilibrium points in the gravitational potential of a uniformly rotating asteroid are the effective potential's critical points (Jiang et al. [8]). Denote $\mathbf{r}_E = (x_E, y_E, z_E)^T$ as the equilibrium point's position vector expressed in the body-fixed frame, $\mathbf{r}$ as the particle's position vector expressed in the body-fixed frame, $\delta \mathbf{r} = [\xi \ \eta \ \zeta]^T = \mathbf{r}_E - \mathbf{r}$ as the particle's position vector viewed from the equilibrium points, $\boldsymbol{\omega} = [\omega_x, \omega_y, \omega_z]^T$, and

$$V_{pq} = \left(\frac{\partial^2 V}{\partial p \partial q}\right)_E \quad (p, q = x, y, z).$$

Then the Taylor expansion for $V(x, y, z)$ at the equilibrium point $\mathbf{r}_E = (x_E, y_E, z_E)^T$ is

$$V(x, y, z) = V(x_E, y_E, z_E) + \frac{1}{2}\left(\frac{\partial^2 V}{\partial x^2}\right)_E (x - x_E)^2 + \frac{1}{2}\left(\frac{\partial^2 V}{\partial y^2}\right)_E (y - y_E)^2 + \frac{1}{2}\left(\frac{\partial^2 V}{\partial z^2}\right)_E (z - z_E)^2$$
$$+ \left(\frac{\partial^2 V}{\partial x \partial y}\right)_E (x - x_E)(y - y_E) + \left(\frac{\partial^2 V}{\partial x \partial z}\right)_E (x - x_E)(z - z_E) + \left(\frac{\partial^2 V}{\partial y \partial z}\right)_E (y - y_E)(z - z_E) + \cdots$$

(30)

Substituting Eq. (30) into Eq. (4) gives

$$\frac{d^2}{dt^2}\begin{bmatrix}\xi\\\eta\\\zeta\end{bmatrix}+\begin{pmatrix}0 & -2\omega & 0\\2\omega & 0 & 0\\0 & 0 & 0\end{pmatrix}\cdot\frac{d}{dt}\begin{bmatrix}\xi\\\eta\\\zeta\end{bmatrix}+\begin{pmatrix}V_{xx} & V_{xy} & V_{xz}\\V_{xy} & V_{yy} & V_{yz}\\V_{xz} & V_{yz} & V_{zz}\end{pmatrix}\cdot\begin{bmatrix}\xi\\\eta\\\zeta\end{bmatrix}=0. \quad (31)$$

Substituting Eq. (30) into Eq. (2) and Eq. (3) gives

$$\frac{d^2}{dt^2}\begin{bmatrix}\xi\\\eta\\\zeta\end{bmatrix}+\begin{pmatrix}0 & -2\omega_z & 2\omega_y\\2\omega_z & 0 & -2\omega_x\\-2\omega_y & 2\omega_x & 0\end{pmatrix}\cdot\frac{d}{dt}\begin{bmatrix}\xi\\\eta\\\zeta\end{bmatrix}+\begin{pmatrix}V_{xx} & V_{xy} & V_{xz}\\V_{xy} & V_{yy} & V_{yz}\\V_{xz} & V_{yz} & V_{zz}\end{pmatrix}\cdot\begin{bmatrix}\xi\\\eta\\\zeta\end{bmatrix}=0. \quad (32)$$

Both of Eq. (31) and Eq.(32) are linearised dynamical equations for the particle relative to the equilibrium points in the gravitational potential of a uniformly rotating asteroid. The difference is that Eq. (31) is for the case of the body-fixed frame chosen such that the unit vector $\mathbf{e}_z$ satisfies $\boldsymbol{\omega}=\omega\mathbf{e}_z$, in other words, Eq. (31) is for the case when z-axis is the spin axis. Eq. (32) is for the case of the body-fixed frame chosen arbitrarily. Using Eq. (31) and Eq.(32), one can derive the characteristic equation of the equilibrium points, which are

$$\begin{vmatrix}\lambda^2+V_{xx} & -2\omega\lambda+V_{xy} & V_{xz}\\2\omega\lambda+V_{xy} & \lambda^2+V_{yy} & V_{yz}\\V_{xz} & V_{yz} & \lambda^2+V_{zz}\end{vmatrix}=0, \quad (33)$$

and

$$\begin{vmatrix}\lambda^2+V_{xx} & -2\omega_z\lambda+V_{xy} & 2\omega_y\lambda+V_{xz}\\2\omega_z\lambda+V_{xy} & \lambda^2+V_{yy} & -2\omega_x\lambda+V_{yz}\\-2\omega_y\lambda+V_{xz} & 2\omega_x\lambda+V_{yz} & \lambda^2+V_{zz}\end{vmatrix}=0. \quad (34)$$

Here Eq. (33) is for Eq. (31) while Eq. (34) is for Eq. (32), $\lambda$ represnets the eigenvalues of the equilibrium point. In this section, we use the case of the body-fixed frame chosen such that the unit vector $\mathbf{e}_z$ satisfies $\boldsymbol{\omega}=\omega\mathbf{e}_z$. Thus we use Eq. (29) to calculate the position vector of the equilibrium points relative to the body-fixed frame of the asteroid, and use Eq. (33) to calculate the eigenvalues of the equilibrium points. The topological cases of the equilibrium points can be classified by the distribution of the eigenvalues. There are several topological cases for the equilibrium points. The most common cases are ordinary case O1, O2, and O4. For the Case O1, the distribution of the eigenvalues is $\pm i\beta_j \left( \beta_j \in \mathrm{R}, \beta_j > 0; j = 1,2,3 | \forall k \neq j, k = 1,2,3, s.t. \beta_k \neq \beta_j \right)$; For the Case O2, the distribution of the eigenvalues is $\pm \alpha_j \left( \alpha_j \in \mathrm{R}, \alpha_j > 0, j = 1 \right)$ and $\pm i\beta_j \left( \beta_j \in \mathrm{R}, \beta_j > 0; j = 1,2 | \beta_1 \neq \beta_2 \right)$; For the Case O4, the distribution of the eigenvalues is $\pm i\beta_j \left( \beta_j \in \mathrm{R}, \beta_j > 0, j = 1 \right)$ and $\pm \sigma \pm i\tau (\sigma, \tau \in \mathrm{R}; \sigma, \tau > 0)$. Case O1 is linearly stable, and both of the Case O2 and O4 are unstable. More detailed topological cases of the equilibrium points can be found in Jiang et al. [30]. One can use equilibria of the large-size-ratio binary asteroid system to derive the equilibria of a single asteroid. In Eq. (19), we don't consider the attitude motion of the primary, then the equilibria of a single asteroid satisfies Eq. (20), which can reduce to Eq. (33).

Figure 4 illustrates the zero-velocity curves and the projections of equilibrium points of asteroid 1333 Cevenola in the xy, yz, and zx planes, respectively. Figure 5 presents effective potential of asteroid 1333 Cevenola in different planes. From the figure, one can see that there are totally five equilibrium points in the potential of 1333 Cevenola. All the five equilibrium points are out-of-plane equilibrium points. In other words, all the five equilibrium points are not in the equatorial plane (i.e. xy plane) of asteroid 1333 Cevenola. However, these five equilibrium points are near the equatorial plane because the asteroid rotates around z-axis. There are three equilibrium points E2, E4, and E5 are near the yz plane. In like manner, there are three equilibrium points E1, E3, and E5 near the zx plane. Yang et al. [3] used the zero-velocity curves in the xy plane of the mass dipole model to analyze the stable regions in the xy plane. In this paper, although there exist stable regions in the xy plane in the vicinity of equilibrium points E2 and E4 around 1333 Cevenola, there are no stable regions in the vicinity of equilibrium points E2 and E4 because these two equilibrium points are unstable. One can also see the zero-velocity curves in the yz plane to find that there exist saddle curves around equilibrium points E2 and E4 in the yz plane. Thus one can conclude that there are no stable regions in the vicinity of equilibrium points E2 and E4 around asteroid 1333 Cevenola.

Table 1 presents the positions, topological cases, as well as the stability of these equilibrium points. The equilibrium points E5 is near the mass center of the body. The equilibrium points E1 and E3 are near the x-axis while E2 and E4 are near the y-axis. Only E5 is inside the body of the asteroid 1333 Cevenola. E5 is stable, other equilibrium points are unstable. The topological cases of the inner equilibrium points E5 is Case O1. The topological cases of the equilibrium points which are near the x-axis belong to Case O2, while topological cases of the equilibrium points which are near the y-axis belong to Case O4. The topological cases of the equilibrium points E1, E2, E3, and E4 are Case O2, Case O4, Case O2, and Case O4. Thus the topological cases of the outside equilibrium points E1, E2, E3, and E4 have a staggered distribution.

Because all the equilibrium points are near the equatorial plane of the asteroid 1333 Cevenola, we can use the structure of the zero-velocity curves and the projections of the equilibrium points to judge the stability of the equilibrium points. From Figure 4(a), one can see that the equilibrium points E1 and E3 are unstable. One cannot judge the stability of the equilibrium points E2, E4, and E5 from Figure 4(a).

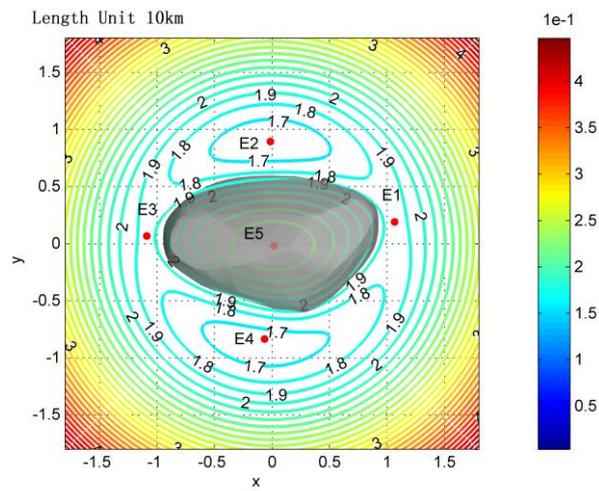

(a)

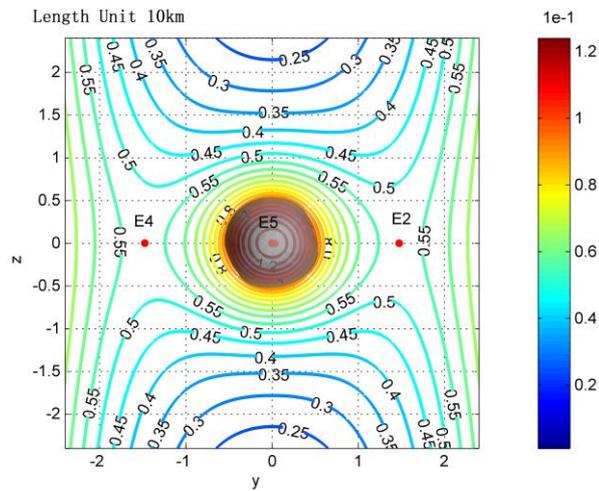

(b)

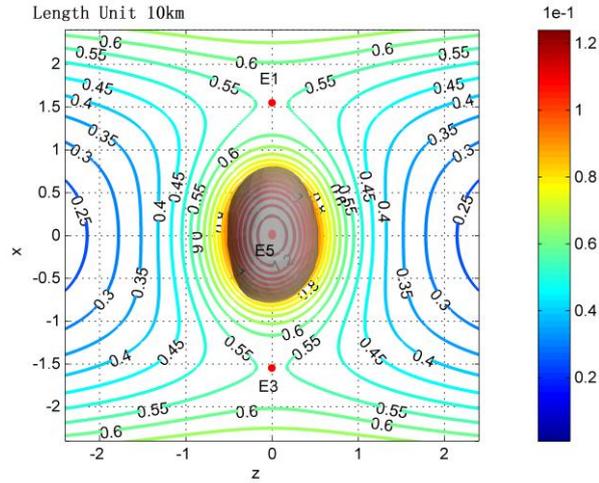

(c)

Figure 4 Zero-velocity curves of asteroid 1333 Cevenola in the xy, yz, and zx planes, respectively. The projections of equilibrium points are also plotted in these three coordinate planes to compare with the zero-velocity curves. The unit of the colour code is m$^2$·s$^{-2}$. (a) The five equilibrium points E1-E5 and the zero-velocity curves of asteroid 1333 Cevenola in the xy plane; (b) The three equilibrium points E2, E4, as well as E5 and the zero-velocity curves of asteroid 1333 Cevenola in the yz plane; (c) The three equilibrium points E1, E3, as well asE5 and the zero-velocity curves of asteroid 1333 Cevenola in the zx plane.

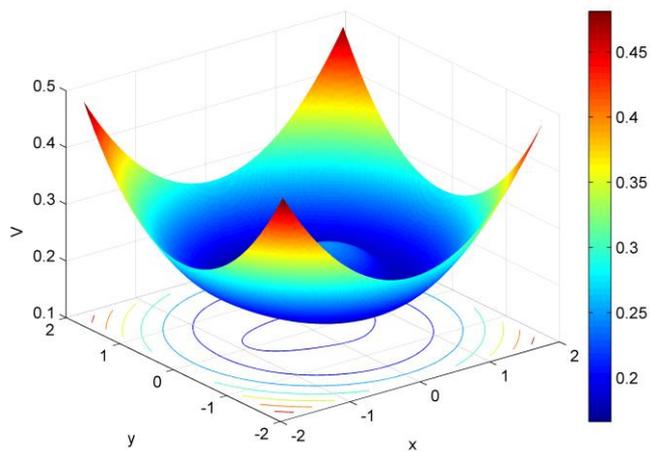

(a)

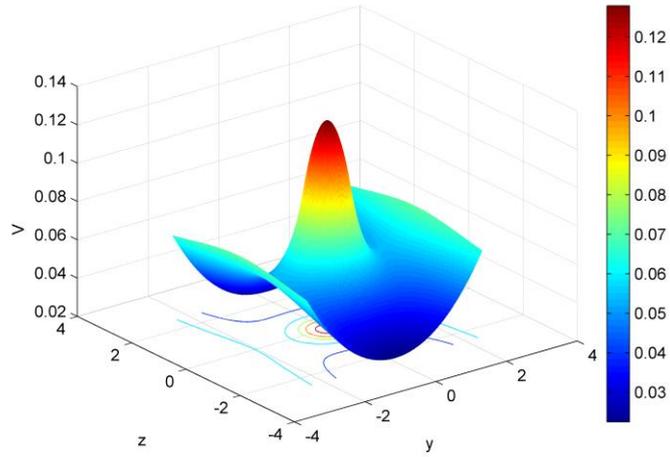

(b)

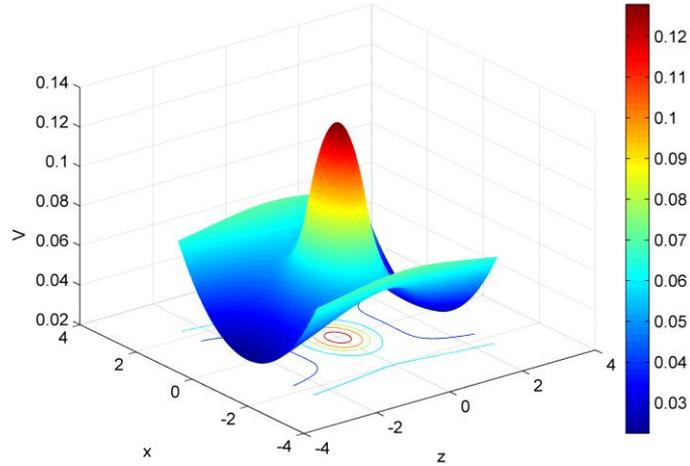

(c)

Figure 5 Effective potential of asteroid 1333 Cevenola in the xy, yz, and zx planes, respectively. The unit of the colour code is $10^4 m^2 \cdot s^{-2}$, and the unit of the length in the coordinate axes x, y, and z is 10km. (a) 3D plot of the effective potential in the xy plane; (b) 3D plot of the effective potential in the yz plane; (c) 3D plot of the effective potential in the zx plane. In each subfigure, the contour of the effective potential is also plotted.

Table 1 Positions, topological cases, and stability of the equilibrium points around 1333 Cevenola

| Equilibrium Points | x (km) | y (km) | z (km) | Cases | Stability |
|---|---|---|---|---|---|
| E1 | 10.6574 | 1.89845 | -0.0438887 | O2 | U |
| E2 | -0.145281 | 8.93534 | -0.0531682 | O4 | U |

| E3 | -10.8971 | 0.662615 | 0.0983730 | O2 | U |
| --- | --- | --- | --- | --- | --- |
| E4 | -0.652500 | -8.35695 | 0.0934736 | O4 | U |
| E5 | 0.184283 | -0.199398 | 0.00696604 | O1 | S |

**4 Orbits Closes to 1333 Cevenola**

In this section, we calculate the 3D orbits closes to 1333 Cevenola to analyze the orbital stability near this main belt binary asteroid. We first calculated an orbit near the equatorial plane with considering the full gravitational potential caused by the irregular shape of 1333 Cevenola. The trajectories are plotted relative to the body-fixed frame and the inertia frame. The mechanical energy and the Jacobian of the orbit are calculated to analyze the dynamical behaviors. The results are showed in Figure 6. The integral time is 351360s. From Figure 6, one can see that the 3D shape of the orbit in the body-fixed frame and the 3D shape of the orbit in the inertia frame are quite different. The amplitude spectrum of the orbit in the body-fixed frame is larger than that in the inertia frame. The mechanical energy varies quasi-periodic. The variation range of the mechanical energy is in the interval [-1.793305617, -1.694391275]J·kg$^{-1}$. The Jacobian of this orbit is a constant and equal -22.2311271317 J·kg$^{-1}$.

To study the motion stability of the inclined orbit closes to 1333 Cevenola, we simulated an inclined orbit in the gravitational potential of 1333 Cevenola with considering the full gravitation caused by the irregular shape. The results are showed in Figure 7. The integral time is 3513600s, ten times than the integral time of the above example. Because the 3D shape of the orbit looks strange, and the result of integral time 351360s cannot bring complete show of the 3D shape. The 3D view of the orbit relative to the body-fixed frame and the inertia frame are both presented. From Figure 7, one can see that the 3D view of the orbit relative to the body-fixed frame repeats every five circles. The whole shape of the orbit in the body-fixed frame looks like a silk ribbon which has five circles. The amplitude spectrum of the orbit in the body-fixed frame is obviously larger than that in the inertia frame. The mechanical energy also varies quasi-periodic. The variation range of the mechanical energy is in the interval [-1.630896473 -1.456852116]J·kg$^{-1}$. The Jacobian of this orbit is a constant and equal -22.0404592288 J·kg$^{-1}$.

Comparing these two orbits, one can see that although the 3D shapes of these two orbits are quite different, the Jacobi integrals of these two orbits are pretty close. The length of the interval for the mechanical energy of the orbit near the equatorial plane is 0.098914342 J·kg$^{-1}$, while the length of the interval for the mechanical energy of the inclined orbit is 0.174044357 J·kg$^{-1}$. The variety of the mechanical energy of the inclined orbit is larger than that of the orbit near the equatorial plane. This is because that the variety of the gravitational environment near the equatorial plane is small. Thus we can conclude that for the moonlet in the gravitational potential of the primary, if the semi-major and the eccentricity are fixed, the moonlet's orbit is more likely to be stable if the orbit is near the equatorial plane of the primary.

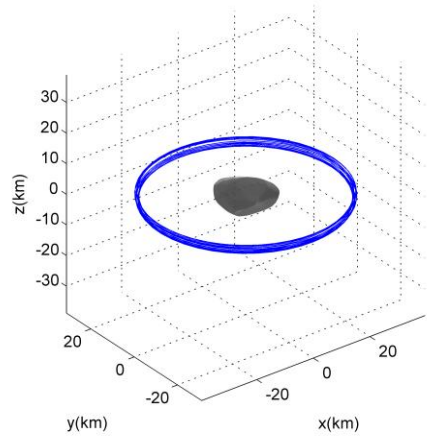

(a)

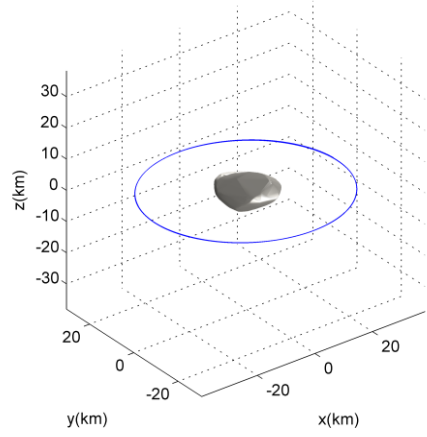

(b)

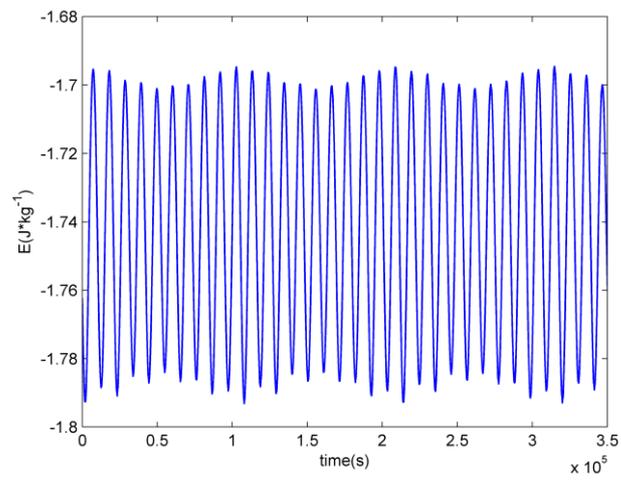

(c)

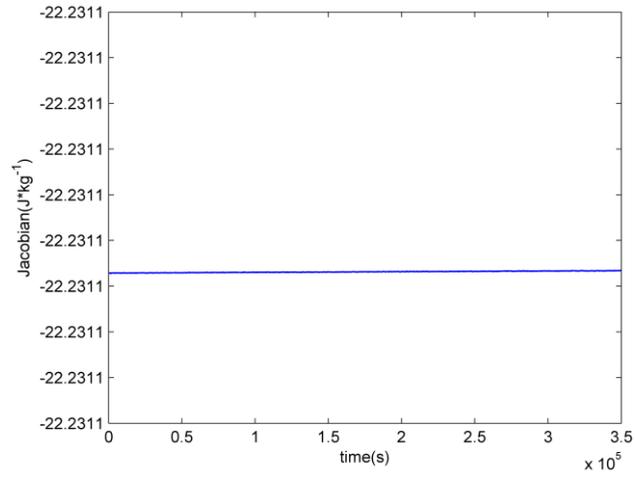

(d)

Figure 6 Examples of 3D equatorial orbit calculated with 351360s. (a) 3D view in the body-fixed frame; (b) 3D view in the inertia frame; (c) The mechanical energy of the orbit; (d) The Jacobian of the orbit.

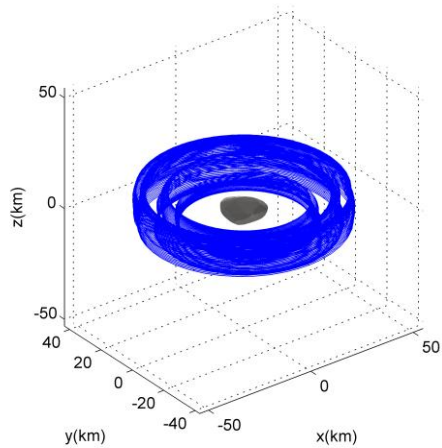

(a)

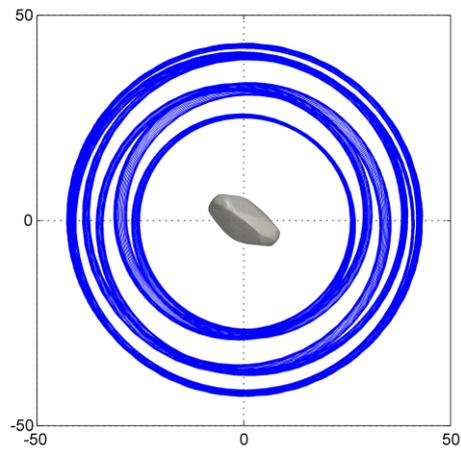

(b)

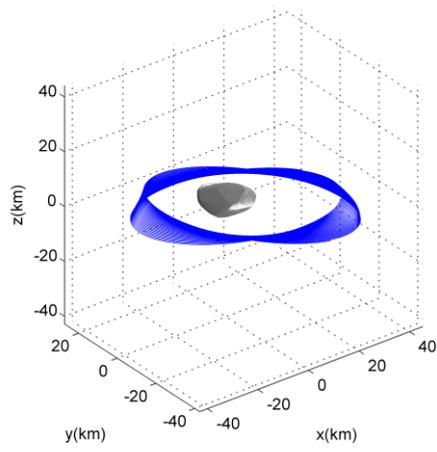

(c)

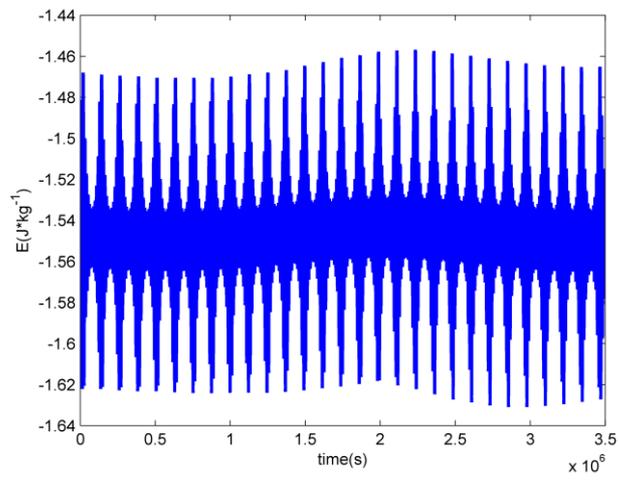

(d)

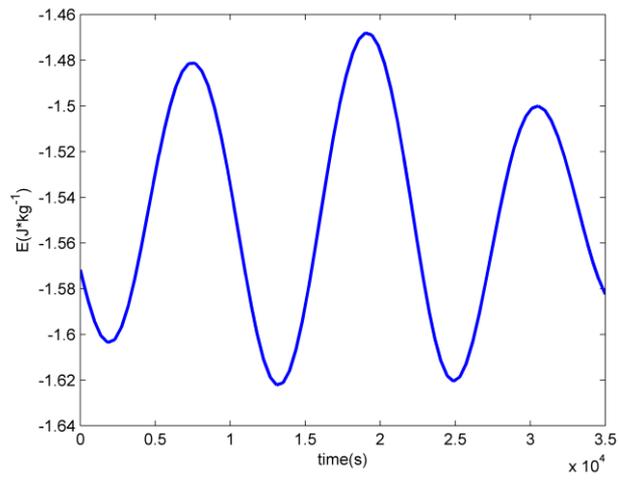

(e)

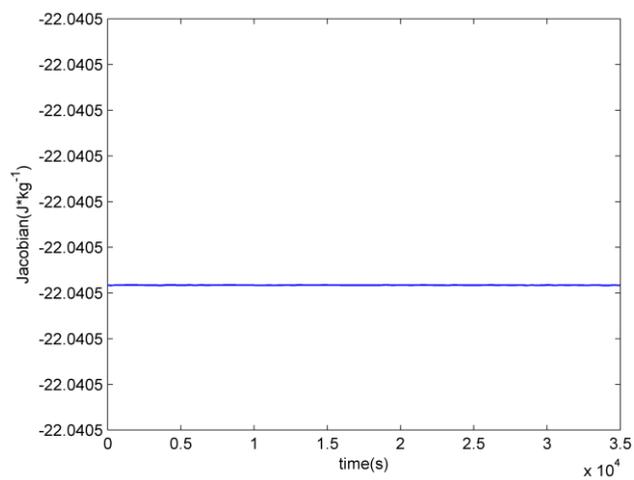

(f)

Figure 7 Examples of 3D inclined orbit calculated with 3513600s. (a) 3D view in the body-fixed frame; (a) Viewed from +z axis; (c) 3D view in the inertia frame; (d) The mechanical energy of the orbit; (e) Local plot of the mechanical energy of the orbit; (f) The Jacobian of the orbit.

## 5 Conclusions

This paper has analyzed the equilibrium points and orbits around asteroid 1333 Cevenola. The 3D irregular shape of the asteroid 1333 Cevenola and the full gravitational potential caused by the 3D irregular shape have been investigated. The zero-velocity curves in the xy, yz, and zx planes are presented. There are five equilibrium points in the gravitational potential of 1333 Cevenola. All the outside equilibrium points are unstable. Two different orbits are calculated with considering the full gravitational potential caused by the 3D irregular shape of 1333 Cevenola. The Jacobi integrals of these two orbits are constants. The result suggests that the moonlet's orbit is more likely to be stable if the orbit inclination is small.

## Acknowledgements

This research was supported by the National Natural Science Foundation of China (No. 11772356) and China Postdoctoral Science Foundation- General Program (No. 2017M610875).